\documentstyle[12pt,axodraw]{article}
\newcommand{\Tint}[1]{{\hbox{$\sum$}\!\!\!\!\!\!\int}_{\!\!\!\!#1}}
\newcommand{\la}[1]{\label{#1}}
\newcommand{\be}{\begin{equation}}
\newcommand{\ee}{\end{equation}}
\newcommand{\ba}{\begin{eqnarray}}
\newcommand{\ea}{\end{eqnarray}}
\newcommand{\bi}{\begin{itemize}}
\newcommand{\ei}{\end{itemize}}

\newcommand{\nr}[1]{(\ref{#1})}
\newcommand{\tr}{{\rm Tr\,}}
\newcommand{\nn}{\nonumber \\}
\newcommand{\fr}[2]{{\frac{#1}{#2}}}

\newcommand{\bfp}{{\bf p}}
\renewcommand{\vec}[1]{{\bf #1}}

\def\lsi{\raise0.3ex\hbox{$<$\kern-0.75em\raise-1.1ex\hbox{$\sim$}}}
\def\gsi{\raise0.3ex\hbox{$>$\kern-0.75em\raise-1.1ex\hbox{$\sim$}}}
\newcommand{\lsim}{\mathop{\lsi}}
\newcommand{\gsim}{\mathop{\gsi}}
\begin{document}

\begin{titlepage}
\begin{flushright}
CERN-TH/97-298\\
NORDITA-97/78P\\
hep-ph/9710538\\
December 1997\\
\end{flushright}
\begin{centering}
\vfill

{\bf HIGH TEMPERATURE DIMENSIONAL REDUCTION\\ AND PARITY VIOLATION}
\vspace{0.8cm}

K. Kajantie$^{\rm a,b}$\footnote{keijo.kajantie@cern.ch},
M. Laine$^{\rm a}$\footnote{mikko.laine@cern.ch},
K. Rummukainen$^{\rm c}$\footnote{kari@nordita.dk} and
M. Shaposhnikov$^{\rm a}$\footnote{mshaposh@nxth04.cern.ch} \\

\vspace{0.3cm}
{\em $^{\rm a}$Theory Division, CERN, CH-1211 Geneva 23,
Switzerland\\}
\vspace{0.3cm}
{\em $^{\rm b}$Department of Physics,
P.O.Box 9, 00014 University of Helsinki, Finland\\}
\vspace{0.3cm}
{\em $^{\rm c}$NORDITA, Blegdamsvej 17,
DK-2100 Copenhagen \O, Denmark}

\vspace{0.7cm}
{\bf Abstract}

\end{centering}

\vspace{0.3cm}\noindent
The effective super-renormalizable 3-dimensional Lagrangian,
describing the high temperature limit of chiral gauge theories, has more
symmetry than the original 4d Lagrangian: parity violation is absent.
Parity violation appears in the 3d theory only through
higher-dimensional operators. We compute the coefficients of dominant
P-odd operators in the Standard Electroweak theory and discuss their
implications. We also clarify the parametric accuracy obtained with
dimensional reduction.
\vfill
\noindent

\end{titlepage}

\section{Introduction}
A general feature of effective field theories is that they may have
more symmetries than the corresponding original theories. These extra
symmetries are then broken by higher-dimensional operators in the
effective theory. In many cases, the higher order operators are
non-renormalizable whereas the ``effective'' Lagrangian is
renormalizable. A familiar example is GUT-induced baryon number
violation~\cite{B+L} which is absent in the renormalizable Standard
Model Lagrangian. Another classic example is the four-Fermion
interaction, which introduces, e.g., strangeness non-conservation into
QED$+$QCD. Additional symmetries may also appear in the case of
non-renormalizable effective theories, like in chiral effective
Lagrangians for QCD \cite{witten}.

The purpose of this paper is to point out that a similar situation may
arise in dimensionally reduced effective field theories describing
the high temperature thermodynamics of various weakly coupled gauge
theories~[3--11] %\cite{dr}--\cite{su5}
%{dr,perturbative,generic,jkp,bn,mssm,su5}
(for a review, see~\cite{erice}). In this context, the effective theory
is 3-dimensional and super-renormalizable. It does not contain parity 
(P) violation, 
nor in many cases, such as the Standard Model, 
charge conjugation (C) violation. 
However, if the original 4d theory
breaks P or C, it will induce P or C breaking higher dimensional
operators into the effective theory.

To be concrete, consider the electroweak sector the Standard Model
(SM), disregarding the U(1) group, at non-zero temperature $T$ but with
zero chemical potentials $\mu=0$ for all conserved charges. The
super-renormalizable effective action relevant for this case
is\footnote{The super-renormalizable effective 3d theory for the case
of
non-zero chemical potentials was discussed in \cite{melting}.}
\ba
S & = & \int\! d^3x \biggl\{
\frac{1}{2}\tr F_{ij}F_{ij}+
(D_i\phi)^{\dagger}(D_i\phi)+
m_3^2\phi^{\dagger}\phi+\lambda_3
(\phi^{\dagger}\phi)^2 \nonumber \\
& & +\tr [D_i,A_0][D_i,A_0]+m_D^2\tr A_0 A_0+
\lambda_A(\tr A_0 A_0)^2
+ 2 h_3\phi^{\dagger}\phi \tr A_0 A_0 \biggr\} ,
\label{action}
\ea
where $F^a_{ij}=\partial_iA_j^a-\partial_jA_i^a+
g_3\epsilon^{abc}A^b_iA^c_j$, $F_{ij}=T^a F^a_{ij}$,
$D_i\phi=(\partial_i-ig_3 T^aA^a_i)\phi$, $A_0=T^a A_0^a$, and
$T^a=\tau^a/2$. The $\tau^a$ are the Pauli matrices. The factor $1/T$
multiplying the action has been scaled into the fields and the coupling
constants, so that the fields have the dimension GeV$^{1/2}$ and the
couplings $g_3^2$, $\lambda_3$ have the dimension GeV. The connection
between the couplings in eq.~\nr{action} and the physical parameters of
the 4d theory is given explicitly in \cite{generic}. The P and C violating
chiral fermions contribute only through the parametric dependence of
the couplings on the number of families
$N_f$ and on the top quark Yukawa coupling
$g_Y$. The terms of lowest dimensionality
neglected in eq.~\nr{action} have dim=6 in 4d units; an example is
$(\phi^\dagger\phi)^3$. The P and C conserving dim=6 operators arising in
dimensional reduction have been studied in
\cite{jkp,chapman,moore}, and P breaking operators have
been discussed in [16--20]. %\cite{bks}--\cite{su2u1}.

It is now obvious that the P and C violating effects of the 4d theory
must appear in the higher dimensional operators of the effective
theory. In fact, one expects that the dominant effects will come from
the integration out of the top quark. The leading diagrams, beyond those
already included in the computation of the coupling constant relations
for the theory in eq.~\nr{action}, are
shown in Fig.~1. The corresponding operator will be given in eq.~\nr{o4}
below.

\begin{figure}[b]

\vspace*{-0.5cm}

\begin{center}
\begin{picture}(300,120)(0,0)

\SetWidth{1.5}
%\Line(40,20)(80,20)
%\Line(80,20)(100,54.5)
%\Line(100,54.5)(80,89)
%\Line(80,89)(40,89)
%\Line(40,89)(20,54.5)
%\Line(20,54.5)(40,20)
\CArc(60,54.5)(40,0,360)
\DashLine(30,3)(40,20){5}
\DashLine(90,3)(80,20){5}
\Photon(30,106)(40,89){1.5}{3}
\Photon(90,106)(80,89){1.5}{3}
\Photon(100,54.5)(120,54.5){1.5}{3}
\Photon(20,54.5)(0,54.5){1.5}{3}

\Text(55,6.5)[l]{$R$}
\Text(100,35)[l]{$L$}
\Text(100,74)[l]{$L$}
\Text(55,102.5)[l]{$L$}
\Text(12,74)[l]{$L$}
\Text(12,35)[l]{$L$}
\Text(-3,56)[r]{$0$}
\Text(25,115)[l]{$i$}
\Text(90,115)[l]{$j$}
\Text(125,56)[l]{$k$}

\CArc(240,54.5)(34.5,0,360)
%\Line(215.6,79.9)(264.4,79.9)
%\Line(215.6,30.1)(264.4,30.1)
%\Line(215.6,30.1)(215.6,79.9)
%\Line(264.4,79.9)(264.4,30.1)
\Photon(263.9,79.4)(284.4,99.9){1.5}{4}
\Photon(216.1,79.4)(195.6,99.9){1.5}{4}
\DashLine(215.6,30.1)(195.6,10.1){5}
\DashLine(264.4,30.1)(284.4,10.1){5}
\Text(190,107)[l]{$0$}
\Text(288,107)[l]{$i$}
\Text(236,97)[l]{$L$}
\Text(194,54)[l]{$L$}
\Text(280,54)[l]{$L$}
\Text(236,12)[l]{$R$}
%\Text(246,90)[l]{$L$}
%\Text(222,54)[l]{$L$}
%\Text(272,54)[l]{$L$}
%\Text(236,20)[l]{$R$}

\end{picture}
\end{center}
\caption[a]{
The graphs giving the dim=6 operator $O_4^{--}$ in eq.~\nr{o4}.
The solid line is a (top) quark propagator, the dashed line a Higgs
field and the wiggly line a gauge field. 
The symbols L and R indicate the
handednesses of the quarks. For the dim=6 operator one needs the
momentum dependence of the 4-leg diagram, which supplies two more
indices $j,k$. An obvious 5-leg diagram is not shown.}
\la{fig1}
\end{figure}
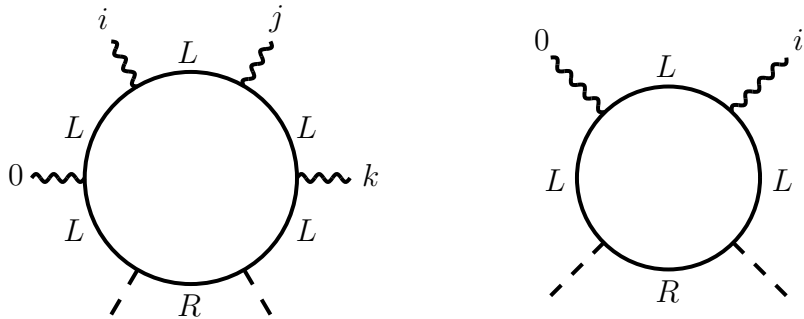

\section{P and C breaking operators}
Before computing the diagram in Fig.~1 it is illuminating to analyse the
situation more generally. To this end we first specify the
transformation properties under discrete 4d CPT transformations of the
fields appearing in the 3d action $S=S[A_i,A_0,\phi]$, then find out
all possible $J^{PC}=0^{+-},0^{-+},0^{--}$ operators of lowest
dimensionalities and finally see which of these are really induced in
the transition from the 4d to the 3d effective theory and with what
coefficients. Since the CP violation of the SM is very small, one
expects the coefficients of any induced $0^{+-},0^{-+}$ operators to be
tiny so that the main interest is in the P and C violating
operators with $J^{PC}=0^{--}$. Note that in some extensions
of the Standard Model, such as in the MSSM, CP-violation
can appear in the bosonic sector even in the super-renormalizable
vertices of the 3d theory, and could thus have correspondingly larger effects. 

The transformation properties of the objects appearing in the effective
theory $S=S[A_i,A_0,\phi]$ under the 4d discrete symmetries C, P and T
are shown in Table~\ref{cpt}. With these one can explicitly verify that
the terms in eq.~\nr{action} are separately invariant under C, P and T.

\begin{table}[b]
\center
\begin{tabular}{|r|r|r|r|r|}
\hline
         &
\multicolumn{1}{|c|}{C}      &
\multicolumn{1}{c|}{P}     &
\multicolumn{1}{c|}{T}     & CPT       \\
\hline
$\phi$   &   $\phi^*$  &   $\phi$   &  $\phi$   & $\phi^*$  \\
$A_0$    & $-A_0^*$    &   $A_0$    &  $A_0$    & $-A_0^*$  \\
$A_i$    &   $-A_i^*$  &   $-A_i$   &  $-A_i$   & $-A_i^*$  \\
$i$      &    $i$      &    $i$     &  $-i$     & $-i$      \\
$D_0$    & $D_0^*$     &   $D_0$    & $-D_0$    & $-D_0^*$  \\
$D_i$    &   $D_i^*$   &   $-D_i$   &  $D_i$    & $-D_i^*$   \\
$B_i$    &   $-B_i^*$   &  $B_i$    &  $-B_i$   & $B_i^*$   \\
$F_{ij}$ & $-F_{ij}^*$ &  $F_{ij}$  & $-F_{ij}$ & $F_{ij}^*$   \\
\hline
\end{tabular}
\caption[a]{The transformation properties of the bosonic fields with
Minkowskian $A_0$. Minkowski space is used here because it is needed
for seeing the Hermiticity properties of the operators. In the text the
operators are written with Euclidian $A_0$, $A_0\equiv A_0^E =
-iA_0^M$, so that they can consistently be added to the
Euclidian action in eq.~\nr{action}. Note that apart from for $\phi$,
$C$ for
SU(2) corresponds to the global gauge transformation $g=i\tau^2$, and
thus there is no $C$-violation without $\phi$. The transformations
could also be written in some other forms; for instance, the
$T$-transformation properties of $A_0^M$, $A_i$ are equivalent to
$iA_0^M\to -(iA_0^M)^*$, $iA_i\to (iA_i)^*$.\la{cpt}}
\end{table}

Consider now the possible P or C violating locally gauge invariant
operators in order of increasing dimensionality.

{\bf 1.} Dim=5. The lowest possible dimensionality is dim=5 in 4d units
(after rescaling to 3d, the dimension is $3\fr12$ or 4). It is helpful
to write the spatial gauge fields in terms of
\be
B_i = \fr12 \epsilon_{ijk} F_{jk}.
\ee
The P violating locally gauge invariant operators
$O^{PC}$ that can in principle emerge, are then 
\ba
O^{-+}_1 & = & i c_1
\phi^\dagger \{D_i, B_i\} \phi, \la{o1} \\
O^{-+}_2 & = & i c_2 \tr [D_i,A_0][B_i,A_0], \\
O^{-+}_3 & = & c_3 \epsilon_{ijk} \tr B_i [D_j,B_k]. \la{o3}
\ea
Here the number of possible operators has been reduced by making use of
the Bianchi identity $[D_i,B_i]=0$, of the antisymmetry in permutations
of the trace $\tr ABC$ for SU(2), and of the property $\tr A [D,B]=-\tr
B[D,A]$. The coefficients $c_i$ are real in order to make the operators
Hermitian (or, as scalars, real) in Minkowski space. The operators in
eqs.~\nr{o1}--\nr{o3} clearly violate P, but using Table~\ref{cpt}, one
can see that they conserve C. Thus they are CP-violating. Moreover,
they conserve T, thus violating CPT (of the original 4d theory). Terms
of this kind can only arise in connection with non-zero chemical
potentials~\cite{bks} which we assumed are zero.

One can also write down P-conserving but C-violating operators,
$J^{PC}=0^{+-}$:
\be
i \phi^\dagger A_0\phi \phi^\dagger\phi ,\qquad
i \phi^\dagger A_0\phi \tr A_0A_0, \la{pm1}
\ee
\be
i \phi^\dagger[D_i, [D_i,A_0]]\phi, \qquad
i \phi^\dagger\{D_i, \{D_i,A_0\}\}\phi. \la{pm2}
\ee
When written in Minkowski-space using $A_0\equiv A_0^E =
-iA_0^M$, these are seen to conserve T 
and thus to violate CPT. Hence the operators in eqs.~\nr{pm1}, \nr{pm2}
can again only arise in connection with non-zero chemical potentials.
The same holds for $J^{PC}=0^{++}$ operators odd in T, such as 
\be
\phi^\dagger\{D_i, [D_i,A_0]\}\phi.
\ee
Note that for zero chemical potentials
there are no regular $J^{PC}=0^{++}$ T-even operators
induced by dimensional reduction at dim=5, either~\cite{moore}.

{\bf 2.} Dim=6. Consider then the next dimension, dim=6 in 4d units
(dim=4, $4\fr12$ or 5 in 3d units). One can find several CP-violating
$J^{PC}=0^{-+}$ operators, for example
\be
i\partial_k(\phi^\dagger\phi)\tr A_0B_k,\qquad
i\partial_k(\tr A_0A_0)\tr A_0B_k,
\ee
\be
\epsilon_{ijk}\tr[D_i,A_0][D_j,A_0][D_k,A_0],\quad
i\epsilon_{ijk}\tr[D_i,[D_j,F_{km}]][D_m,A_0].
\ee
When written in Minkowski-space, one can see that these operators are real,
have T=$-1$, and hence have CPT=+1. Thus they can appear in the
effective Lagrangian with very small coefficients coming from 
the CP-violation in the original theory. In the MSSM with
new sources of CP-violation and more bosonic fields, 
many other kinds of CP-violating operators can arise, as well. 

On the other hand, there are at dim=6 also
the following two independent CP-even operators:
\ba
O^{--}_4 & = &  c_4 \phi^\dagger \{D_i,\{A_0,B_i\}\} \phi \nn
& =&
2 i c_4 {\rm Im}[\phi^\dagger D_i \phi]
\tr A_0 B_i, \la{o4} \\
O^{--}_5 & = &  c_5
\phi^\dagger [D_i,[A_0,B_i]] \phi. \la{o5}
\ea
These operators are even in T, thus again conserving CPT. The two forms of
$O^{--}_4$ in eq.~\nr{o4}
differ by a 3-divergence for SU(2), and the second form states
explicitly that $O^{--}_4$ is purely imaginary
in Euclidian space for real $c_4$. Since $O^{--}_4$, $O^{--}_5$
conserve CP, the coefficients need not be vanishingly small.

The operators in eqs.~(\ref{o4}),(\ref{o5})
can clearly come from the diagrams in
Fig.~1. To compute the operator corresponding to these diagrams it is
simplest to take the 6-leg diagram since then the external legs can be
taken at zero momentum. If $p$ is the loop momentum and the external
vector legs have the indices $\mu\nu\alpha\beta$, then the
$\gamma_5$-part
of the trace over the fermion loop has the structure
\be
-2p_\gamma(p_\nu\epsilon_{\mu\gamma\alpha\beta}+p_\beta
\epsilon_{\mu\nu\alpha\gamma})+p^2\epsilon_{\mu\nu\alpha\beta}.
\ee
This is now summed over $p_0=(2n+1)\pi T$ and integrated over $\bfp$
by first using $p_0p_i\to0$, $p_ip_j\to\fr13\bfp^2\delta_{ij}$. One
then obtains the structure
\be
\epsilon_{ijk}\phi^\dagger(A_0A_iA_jA_k+A_iA_0A_jA_k+A_iA_jA_0A_k+
A_iA_jA_kA_0)\phi.\la{aaaa}
\ee
Reintroducing $D_i$ and $F_{ij}$ gives precisely the structure
in eq.~\nr{o4},
so that to 1-loop order,
\ba
c_4 & = & \fr23 g^2 g_Y^2 T D_6, \\
c_5 & = & 0,
\ea
where the extra $T$ comes
from going into 3d units and
\be
D_6=\Tint{p_f}\frac{1}{(p^2)^3}=
T\sum_n\int\frac{d^3p}{(2\pi)^3}\frac{1}{[\vec{p}^2+
(2n+1)^2 (\pi T)^2]^3}=\frac{7\zeta(3)}{128\pi^4T^2}.
\ee
The structure in eq.~\nr{o5} would have been obtained if the signs of
the
first and last terms in eq.~\nr{aaaa} had been opposite.

We thus find that to leading order the inclusion of parity violating
effects to the 3d effective theory of finite 
temperature SM leads to a purely
imaginary higher dimensional term in the Euclidian
action. The reason is actually
simple: in Minkowski space the action corresponding to the diagrams in
Fig.~1 must be real. Since the action is linear in $A_0$, the transition
to Euclidian space brings in one imaginary unit $i$.

The fact that the Euclidian P-violating operator $O_4^{--}$
is imaginary, also implies that 
lattice Monte Carlo simulations of P-violating effects 
are in practice very difficult. 
%which can be reduced to expectation values of an even number of
%parity violating operators in the leading super-renormalizable theory. 
Nevertheless, the new terms can have significant effects,
as discussed below.

Note that both of the operators in eqs.~\nr{o4}, \nr{o5} contain the
field $A_0$. On the other hand, around the electroweak phase transition
temperature, the field $A_0$ can be integrated out. This is because
$A_0$ has a mass parameter which is always parametrically of order
$m_D^2\sim g^2T^2$. In contrast, the mass parameter of the scalar field
is $m_3^2\sim g^4T^2$ around the critical temperature, since there the
1-loop term $\sim g^2T^2$ cancels against the tree-level 
term $\sim -m_H^2$. Thus near
$T=T_c$, $A_0$ is ``heavy'' and $\phi, A_i$ are ``light'' and $A_0$ can
be integrated out. Above the critical temperature, both $A_0$ and
$\phi$ are ``heavy'' and can either both be kept in the effective
theory or both be integrated out. In the theory without $A_0$, the
operators $O^{--}_4$, $O^{--}_5$ do not exist and one has to go to
still higher dimensions.

{\bf 3.} Dim=7. We are now interested in P and C violating but CP-even
operators which do not contain $A_0$ and which could hence appear in
the effective theory with only $A_i,\phi$. The following
operators exist at dim=7:
\ba
O^{--}_6 & = &  c_6 \phi^\dagger B_i \phi
\partial_i\phi^\dagger\phi,  \la{o6} \\
O^{--}_7 & = &  i c_7
\epsilon_{ijk} \phi^\dagger [D_i,[B_j,B_k]] \phi, \\
O^{--}_8 & = &  c_8
\phi^\dagger \{[D_i,B_j],\{D_i,D_j\}\} \phi, \\
O^{--}_9 & = &  c_9
\phi^\dagger [\{D_i,B_j\},\{D_i,D_j\}] \phi. \la{o9}
\ea
There cannot be an $O^{--}$ operator without $\phi$, since for SU(2)
C corresponds just to a gauge transformation for all fields but $\phi$
(see Table~1) and the Lagrangian must be invariant under it.

However, even though one can write down the terms in
eqs.~\nr{o6}-\nr{o9} in 3d, one cannot obtain such operators with a
dimensional reduction computation from 4d with zero chemical
potentials.
This is because all these
terms are T-odd and hence CPT-odd (in the 4d theory) according to
Table~1. The statement that the P-violating operators without $A_0$ are
CPT-odd is even more general: it can be seen from Table~1 that in the
absence of $A_0$, CT corresponds to complex conjugation so that any
Hermitian (real) P-odd operator is odd in CPT. 
Thus for $\mu=0$, finite temperature
parity violation can only appear in the 3d effective theory
of eq.~\nr{action} including $A_0$. 

\section{Consequences of the P and C violating operators}
Let us now consider some implications of the operators
discussed. We have three comments to make.

{\bf 1.} The first implication concerns the general statement
of dimensional reduction. According to the arguments in the Appendix,
the following statement can be made:

Consider bosonic static $n$-point one-particle-irreducible (1PI)
Matsubara Green's functions $G^{(4)}_{n}(\vec{p}_i)$ 
for the light ($m\sim g^2T$) and heavy ($m\sim gT$) fields 
in the full 4d theory, depending on external 3-momenta
$\vec{p}_i$. On the other hand, consider the corresponding 1PI Green's
functions in the 3d theory of eq.~\nr{action}. Multiply
the 4d Green's functions by the factor $T^{n/2-1}$ to
have the same dimension GeV$^{3-n/2}$ as the 3d Green's functions, 
and take a region of
temperatures where $|m_3^2| \gsim {\cal O}(g^4 T^2)$. 
Then, there is a mapping of
the temperature and the 4d coupling constants of the underlying theory
to the 3d theory in eq.~\nr{action} such that the 3d theory gives the
same light and heavy parity conserving Green's functions as the full 4d
theory for $p\le gT$ with a relative error at most 
of the order ${\cal O}(g^3)$,
\be
\frac{\Delta G}{G} \lsim {\cal O}(g^3).
\label{accur}
\ee
The error arises from a powercounting estimate of the contributions of
the neglected higher-dimensional operators to typical Green's functions
inside the effective theory (see the Appendix). 
A similar conjecture was first made in Ref.~\cite{generic} but with a
more optimistic error estimate $\Delta G/{G} \sim {\cal O}(g^4)$.   
Let us here clarify a few points related to eq.~\nr{accur}.

First, note that 
near the critical point of the electroweak theory where the line of
first order phase transition ends~\cite{nonpert}, 
the {\em relative} accuracy of
Green's functions determination actually
decreases because of the following
obvious reason. Suppose that the parameters of the effective 3d theory
are such that we are precisely at the critical point. Then the scalar
mass, defined by the two-point Green's function, is exactly equal to zero. 
At the same time, there can be a mismatch in the parameters
of the theory of the relative order $g^3$, so that in the full 4d theory, 
the same Green's function need not be exactly zero for precisely the same
Higgs mass and temperature. In this sense, 
the relative accuracy ${\cal O}(g^3)$ is lost at this
special point. Of course the endpoint itself exists also in the 
full 4d theory, but the corresponding Higgs mass value
$m_{H,c}$ is displaced by a relative  
amount at most of order ${\cal O}(g^3)$. 
 
Second, note that the statement 
concerning the accuracy 
of dimensional reduction in
eq.~\nr{accur} clearly does not apply to P-violating Green's functions.
Indeed, these are exactly zero within the effective theory of
eq.~\nr{action}, yet non-zero in the full theory. 
The statement in eq.~\nr{accur} only applies to P-even 
1PI Green's functions which are non-zero in the effective theory.

It is important to stress that the Green's functions appearing in the
conjecture of dimensional reduction are 1PI. For example, a correlation
matrix employed in computing the masses of excitations coupling to each
other, does not belong to this class. This case is discussed in more
detail in point 2 below.

Finally, it should be noted that observables such as the critical
temperature, the surface tension, or the latent heat, can be determined
from P-even quantities and thus the super-renormalizable Lagrangian in
eq.~\nr{action} is sufficient for all practical purposes. In
particular, non-perturbative results on these observables 
such as in~\cite{nonpert} remain intact by an
addition of P-violating operators.

{\bf 2.} Consider the matrix of two-point correlators defined
by the following commonly used operators:
\ba
H  & = & \phi^\dagger\phi, \hspace*{3.8cm} ({\rm J^{PC}}=0^{++}) \nn
W^3_i & = & i\left[(D_i\phi)^\dagger\phi-\phi^\dagger D_i\phi\right],
\quad\quad ({\rm J^{PC}}=1^{--}) \nn
h_i & = & \tr A_0 B_i.
\hspace*{3.15cm} ({\rm J^{PC}}=1^{++})
\ea
Note that by acquiring a non-zero momentum, the scalar state $H$ can
couple to an operator with quantum numbers ${\rm J^{PC}}=1^{-+}$.

Now, in the super-renormalizable theory, the quantum numbers P and C
are conserved: all the operators in the Lagrangian are $0^{++}$. This
means that the states above cannot couple to each other, and their
correlation matrix would only have diagonal components. However, in the
full theory, there is P-violation, manifesting itself for instance
through the existence of the $0^{--}$-operator in eq.~\nr{o4}
which has a single $A_0$. This
means that states with $J^{PC}=1^{++}$ and $1^{--}$ can couple to
each other, and there are therefore non-zero cross-correlations between
$W^3_i$ and $h_i$. The very long distance exponential fall-off of both
operators is thus determined by a common mass, in contrast to what the
super-renormalizable theory would suggest. This general problem has been
discussed in~\cite{ay}. Thus for this kind of Green's
functions, the effects of P-violation 
are qualitatively important.

{\bf 3.} The third implication of the explicit parity violation induced
by the higher order operators is related to the possibility of
spontaneous parity violation. A priori, it is not excluded that the
theory in eq.~\nr{action} exhibit spontaneous parity breaking in the
symmetric phase \cite{versus}. However, lattice simulations in the
symmetric phase of the SU(2)+Higgs and SU(2)$\times$U(1)+Higgs models
carried out in \cite{parlatt,su2u1} did not show any signal of
spontaneous parity breaking. If such a phenomenon had taken
place, then the small explicit P-violating
operators in eqs.~\nr{o1}-\nr{o3} induced
by chemical potential could have
altered the ground state of the electroweak theory significantly.
Note that, in contrast, the operator in eq.~\nr{o4} cannot have
such an effect, as it is purely imaginary in Euclidian space~\cite{vw}.

\section{Conclusions}

We have discussed the role of parity violation at finite
temperature. For most practical purposes, in particular for the
determination of the thermodynamical properties of the electroweak
phase transition, parity violation is not important, apart from
changing the number of fermionic degrees of freedom participating in
weak interactions. However, there are circumstances, such as the
measurement of the Debye mass in the electroweak theory~\cite{ay}, 
where explicit parity violation does play a role.

\appendix
\renewcommand{\thesection}{Appendix} %~\Alph{section}}

\section{}

In this appendix we give the 
power counting argument behind the
conjecture in eq.~\nr{accur}. 

Let us first recall that slightly different formulations have been
considered for dimensional reduction in the literature. The original
one is a direct integration over the non-zero Matsubara modes (see,
e.g.,~\cite{dr,jkp}). However, at higher than 1-loop level, one has to
be careful with this formulation, and, for instance, it turns out that
in order to construct a local effective theory it is necessary to
consider also the zero Matsubara modes in some internal loops in the
reduction step~\cite{perturbative,generic,jakovac}. The other
formulation is based on matching~\cite{generic,bn}, and in this
formulation the complication mentioned is automatically taken care of.

To now consider the accuracy of the matching procedure, let us assume
the power counting rules $g_Y \sim g$, $\lambda \sim g^2$
characteristic of the renormalization structure of the theory. We also
assume that the exact results of the full 4d theory could be derived
from an effective 3d theory containing an infinite number of higher
order operators, obtained by dimensional reduction. The fields of this
effective theory are generically denoted by $\phi$. We assume that
renormalization is taken care of by the appropriate counter\-terms so
that the only momentum scales in the effective theory are $gT$, $g^2T$.
The question is now that if the 3d theory is truncated such that only
the super-renormalizable Lagrangian in eq.~\nr{action} remains, what is
the error induced for an arbitrary static Green's function? Note that to
minimize the error, one is allowed to make an optimal choice for the
values of the parameters remaining in the effective theory.

In general, the higher order operators generated by 1-loop dimensional
reduction and removed by the truncation, 
are in momentum space of the form 
\be 
{O}_{4d} = g^n\frac{1}{T^{D-4}}\phi^n p^{D-n}, \la{ops4d} 
\ee 
where $D\ge 6$ is the dimension of the operator in 4d units, 
$2\le n\le D$, and $p$ is a generic momentum. 
If some operator is generated only at higher than 1-loop
level, then the coupling constant appears in a still higher power. When
one goes to 3d units, there is one extra $1/T$ multiplying the
Lagrangian and at the same time the fields are scaled by $T^{1/2}$ and
the couplings by its inverse, so that the operators in eq.~\nr{ops4d}
become
\be
{O}_{3d} = g_3^n\frac{1}{T^{D-3}}\phi_{3}^n p^{D-n}. \la{ops3d}
\ee

We now study non-vanishing P-even 1PI Green's function inside the
effective theory of eq.~\nr{action}, and ask what kind of corrections
arise from operators of the type in eq.~\nr{ops3d}. The relevance of
the 1PI-condition is that all the loop momenta are of order $gT, g^2T$.
Otherwise one could assign an arbitrarily small momentum to one of the
lines.

One way of verifying eq.~\nr{accur} is to apply
Symanzik's conjecture \cite{sym} for effective field theories,
which is widely used in the construction of improved 
lattice actions (see, e.g., \cite{lush}). In the present context, 
the conjecture can be viewed as basically a dimensional one 
and it becomes applicable when, after rescalings into the form in 
eq.~\nr{ops3d}, the inverse temperature $1/T$ is (dimensionally) 
identified with the lattice spacing $a$. It is then important to 
note that as has been discussed in the text (see also~\cite{moore}
for the P-even operators), the higher order operators start 
at $D=6$ in the absence of chemical potentials.
Then, the higher dimensional operators in eq.~\nr{ops3d}
are at most of order ${\cal O}(a^3)$. 
Adopting Symanzik's statement to our case, one then knows that 
in order to reproduce effects of order ${\cal O}(a^3)$, one should add 
to the super-renormalizable theory in eq.~(\ref{action}) 
all operators with $D=6$, and
modify the parameters of this theory (couplings, masses, and wave
function normalizations), adding to the tree-level terms 
finite scaling corrections of order ${\cal O}(a^n)$ up to $n=3$. 
Due to the fact that the effective 3d
theory is super-renormalizable and contains dimensionful couplings only,
the mapping can be done perturbatively by computing a finite
number of multiloop diagrams (this is different in 4d at zero 
temperature, where the couplings
are dimensionless and the mapping should be done in a non-perturbative
way, see \cite{lush}). If this is done, then the errors remaining
are ${\cal O}(a/\xi)^4\sim {\cal O}(g^4)$, where the shortest physical 
correlation length $\xi$ of the effective theory
is of the order of the inverse Debye mass, $\xi\sim m_D^{-1}$.

If, on the other hand, the operators with $D=6$ are {\em not
included}, then the effective super-renormalizable theory can 
in principle
have relative errors 
as large as ${\cal O}(a/\xi)^3\sim {\cal O}(g^3)$, 
as stated in eq.~\nr{accur}. To be consistent with 
this precision, it is enough 
to compute the 3d parameters with the relative accuracy $g^2$.
For the coupling constants and wave function normalizations, 
one only needs a  1-loop computation, but for the scalar
mass parameter one needs to go to 3-loop level~\cite{generic} 
(or 2-loop level
if the mass parameter is assumed to be 
of order $\sim g^2T^2$).

Finally, let us note that the above arguments can be directly repeated
for the ``second step'' of dimensional reduction, i.e., for the
integration out of $A_0$. One can see that the relative errors are
${\cal O}(g_3^2 a_0)^3\sim {\cal O}(g^3)$ as stated in~\cite{generic},
where the cutoff scale $1/a_0$ is now identified with the Debye mass
$m_D\sim g T$, and the only momentum scale in the final effective
theory is $\sim g_3^2$. 
As mentioned above, P-violation is non-existent in
the theory without $A_0$, so that it does not cause any complications.

It is interesting to note that, in general, the relative accuracy obtained
with a theory where $A_0$ is kept, is not better than with a theory from 
where $A_0$ is integrated out. Of course, the sets of Green's functions 
considered are different. For those 
Green's functions which are considered in the theory without $A_0$,
the theory with $A_0$ is in principle more accurate.

\end{document}